\documentclass[smallextended]{svjour3}      % onecolumn (second format)
\usepackage{graphicx}
\usepackage{mathptmx}      % use Times fonts if available on your TeX system
\usepackage{amsmath,amssymb}
\journalname{Journal of Visualization}
\begin{document}

\title{Dual-camera system \\for high-speed imaging in particle image velocimetry}

\titlerunning{Dual-camera system for PIV}

\author{K.~Hashimoto \and
        A.~Hori      \and
        T.~Hara      \and
        S.~Onogi     \and
        H.~Mouri
}


\institute{K.~Hashimoto \and A.~Hori \and T.~Hara
              \at 
              Meteorological and Environmental Sensing Technology, Inc., Nanpeidai, Ami 300-0312, Japan \\
           \and
           S.~Onogi \and H.~Mouri (corresponding author)
              \at
              Meteorological Research Institute, Nagamine, Tsukuba 305-0052, Japan \\
              \email{hmouri@mri-jma.go.jp}
}

\date{Received: 2 September 2011 / Accepted: 14 February 2012}

\maketitle

\section{Introduction} \label{s1}

Particle image velocimetry (PIV) is an important technique in experimental fluid mechanics to obtain instantaneous fields of the velocity vectors. The fluid is seeded with tracer particles that follow the flow, which are illuminated by a sheet of light. By using a high-speed camera, two successive images of the particles are taken. The displacements of the particles between the two images determine the velocity vectors of the flow (for details, see Raffel et al. 2007 \cite{r07}; Adrian and Westerweel 2010 \cite{aw10}).

The interval between times for the two images has to be short, especially when the flow is fast. It has been essential to prepare a high-speed camera. However, the high speed is achieved at the expense of other performances of the camera. Even at the time of writing this paper, the sensitivity remains $\lesssim$ ISO $2000$ in the color mode and the sensor resolution remains $\lesssim 4000 \times 3000$. These performances are significantly lower than those of top-end still cameras. Although high-speed cameras are being improved, such a lag behind still cameras does not vanish because they are being improved as well. The high speed also raises the cost of the camera, which prevents some researchers from PIV. 

We point out that the two images could be taken individually with two still cameras. Even if the cameras are those at low cost, they could achieve the same performance as a standard high-speed camera. More importantly, there is a possibility to be able to make an imaging system that has a higher sensitivity or a higher resolution than any high-speed camera. The idea is simple and is probably not new, but we have found no example that such a system was actually made and used for PIV. This paper is the first demonstration of PIV with two still cameras. The imaging system is described in \S\ref{s2}. An example of the PIV is described in \S\ref{s3}. The remarks are summarized in \S\ref{s4}.

%_______________________________________
\begin{figure}[t]
\resizebox{11.5cm}{!}{\includegraphics*[0.5cm,10cm][21cm,20cm]{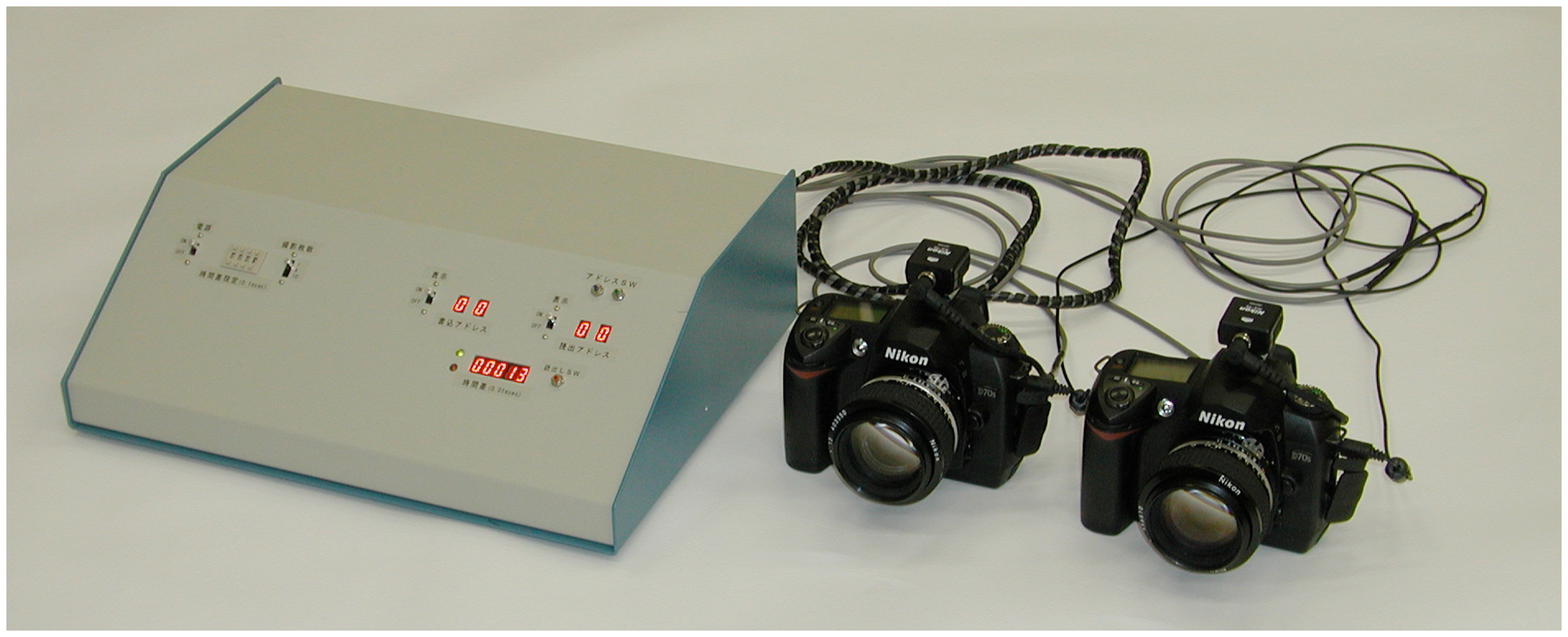}}
\caption{Photo of our imaging system composed of two cameras and an electronic controller}
\label{f1}      
\end{figure}
%_______________________________________

\section{System configuration} \label{s2}

Figure \ref{f1} shows our prototype of the imaging system, which consists of two standard digital still cameras (Nikon, D70S) and an electronic controller. The cameras had $23.7 \times 15.6$\,mm$^2$ CCDs with $3008 \times 2000$ pixels and the sensitivity of ISO $1600$. To trigger their shutters, the controller sends a signal to the one camera and then to the other camera after an interval that is able to be set at any value. In this way, we take two successive images.

Since ours are single-lens reflex cameras with a mechanical mirror and shutter system, there is a lag between triggering the shutter and actually taking the image. This shutter lag is not negligible and is not constant even if the mirror could be locked up. To measure the actual interval between times for the two images, our controller records signals for flash synchronization from both the cameras. 

The focuses of the two cameras are to be on the same area of the light sheet. While the one camera is at the front of the area, the other camera is at the right or left side. Their images are inclined with each other. To correct for this inclination, we take the images of the same reference pattern and estimate the displacement vector at each point of the area. Since the estimation is not free from uncertainty (see below), we repeat this process for different patterns. The mean displacement vector is to be subtracted from the velocity vector.

%_______________________________________
\begin{figure}[tbp]
\resizebox{10.5cm}{!}{\includegraphics*[2.0cm,3.5cm][30.0cm,22.5cm]{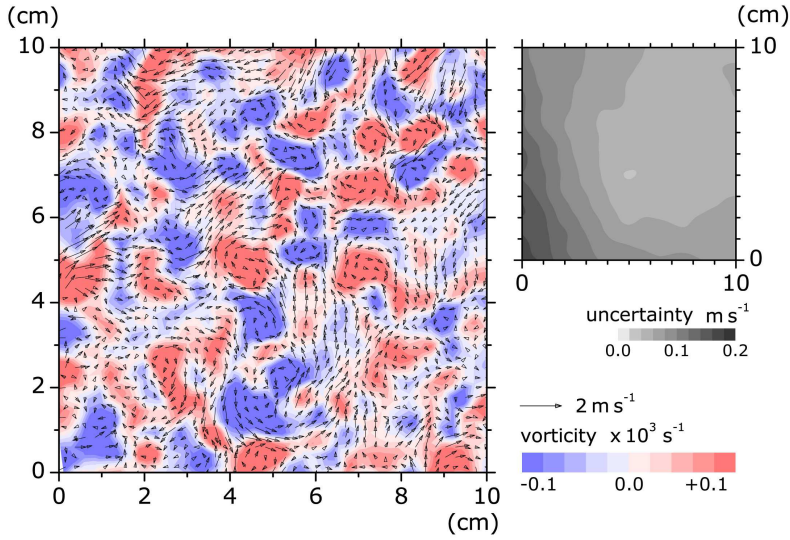}}
\caption{{\it Left:} contour plot of vorticity overlaid with the velocity vector map, with the horizontal axis in the mean flow direction, {\it Right:} contour plot of uncertainty in units of the corresponding velocity}
\label{f2}      
\end{figure}
%_______________________________________

\section{Demonstration} \label{s3}

To demonstrate that our imaging system works equally well as a standard high-speed camera, we present an example, i.e., two-dimensional PIV of grid turbulence in a small wind tunnel of the Meteorological Research Institute (see Mouri et al. 2002 \cite{m02}).

Across the test section of the wind tunnel, which had the size of $0.8 \times 0.8 \times 3$\,m$^3$, we placed a grid made of rods with $0.5 \times 0.5$\,cm$^2$ cross section and with $2.0$\,cm spacing. The mean wind speed was set at $8$\,m\,s$^{-1}$. At about $50$\,cm downstream of the grid, where the turbulence had become fully developed, we obtained an instantaneous field of the velocity vectors. 

We used oil droplets as the tracer particles (Dantec, Super Fog Fluid). The sheet of light with $1$\,mm thickness was provided from an optics with double-pulse Nd:YAG laser (New Wave Research, Solo III-15). The pulse interval was set at $0.25$\,ms. To synchronize the laser pulses with the camera shutters, our controller was equipped with an electronic circuit that triggers the first of the two pulses in response to the signal for flash synchronization from the camera for the first of the two images.

We placed the two cameras at a distance of $70$\,cm from the light sheet. They had $30 \times 20$\,cm$^2$ fields of view practically in focus, through F1.2 lenses with $50$\,mm focal length. Although the lens centers were separated by $17$\,cm, $90$\,\% of their fields were overlapping. Within them,  over an area of about $15 \times 15$\,cm$^2$, the tracer particles were illuminated enough to be discernible.

The images were analyzed with the correlation method to estimate the displacements of the tracer particles (see Raffel et al. 2007 \cite{r07}; Adrian and Westerweel 2010 \cite{aw10}). Then, we corrected for the relative inclination of the two cameras. The images of the reference patterns used here were those of the tracer particles that had been taken at the same times, which were also analyzed with the correlation method.

Figure \ref{f2} shows the contour plot of vorticity overlaid with the velocity vector map (left), from which we have removed the mean flow. To indicate the uncertainty, also shown is the contour plot of standard deviation of the estimated lengths of the displacement vector among different pairs of the same-time particle images (right). Its average corresponds to the velocity of 0.09\,m\,s$^{-1}$. This is well below the average of the turbulence velocity, $0.44$\,m\,s$^{-1}$. Observed in Fig. \ref{f2} are eddies with circulating motion and with enhanced vorticity. Since their sizes are comparable to the spacing of the grid rods, it follows that our imaging system has successfully captured the so-called energy-containing eddies. We have also ascertained that the result is almost of the same quality as with a high-speed camera specialized to PIV.

\section{Concluding remarks} \label{s4}

Having demonstrated that PIV is possible with two still cameras, we expect some applications. In particular, if we use top-end still cameras that are of high sensitivity or of high resolution, the PIV could have a high performance that has never been achieved so far. This could lead to a new finding or a new insight. If we instead use standard still cameras that are of relatively low cost, we could reduce the total cost of PIV without reducing its performance from the standard level. The followings are remarks for these and other applications.

The sensors of our cameras are CCDs, which take the entire image at exactly the same time. This global shutter is just suited to PIV. On the other hand, usually used at present in standard digital still cameras are CMOS sensors. Most of them scan the image, all parts of which are not taken at exactly the same time. There still exist CMOS sensors with global shutters. Such sensors are expected to become usual in digital still cameras.

The shutter lag reduces the efficiency of the imaging, especially when the lag is large and is not constant. For example, in our measurement described above,  the lag had the standard deviation of about 1\,ms. Then, 90\% of the pairs of the images were out of timing with those of the laser pulses. Nevertheless, the controller of our system is readily adapted to mirror-less interchangeable lens cameras that are now becoming standard. Their shutter lag is expected to be short and constant if the shutter is not mechanical but is entirely electronic.

To correct for the relative inclination of the images of the two cameras, we have post-processed the velocity vectors. It would be better to split the same image into the two cameras by using, e.g., a half-silvered mirror, if the cameras are to be placed near the light sheet or if the light sheet is to be thick, although these are rather special cases.

\begin{acknowledgements}
This work was supported in part by KAKENHI Grant No. 22540402.
\end{acknowledgements}


\begin{thebibliography}{}

\bibitem{aw10} Adrian RJ, Westerweel J (2010) Particle Image Velocimetry. Cambridge University Press, Cambridge

\bibitem{m02} Mouri H, Takaoka M, Hori A, Kawashima Y (2002) Probability density function of turbulent velocity fluctuations. Phys Rev E 65: 056304
\bibitem{r07} Raffel M, Willert CE, Wereley ST, Kompenhans J (2007) Particle Image Velocimetry: a Practical Guide, 2nd edition. Springer, Berlin

% etc
\end{thebibliography}
\end{document}